\documentstyle[aps,prb,multicol,epsf]{revtex}
\begin{document}
\title
{Effect of van Hove Singularities on a Spin Liquid}
\author{L. B. Ioffe}
\address{Physics Department \\
Rutgers University, Piscataway, NJ 08855 \\
and \\
Landau Institute for Theoretical Physics, Moscow}
\author{A. J. Millis}
\address{AT\&T Bell Laboratories \\
600 Mountain Avenue  \\
Murray Hill, NJ 07974 }
\maketitle

\begin{abstract}
We determine the properties and leading instabilities of a spin liquid with a
Fermi surface passing near a van Hove singularity.
Our study is motivated by recent photoemission experiments on high $T_c$
cuprates in which it is found that for the optimally doped material
the experimental Fermi surface passes near a van Hove singularity, while for
underdoped materials, a pseudogap in the electron spectral function is formed
in the vicinity of the van Hove point.
We show theoretically that proximity to the van Hove singularity suppresses
the inelastic scattering due to the gauge field and permits the formation of a
d-wave RVB state in which the gap exists only near the van Hove points while
finite regions of the Fermi surface remain gapless.
This $d$-wave pairing provides a natural explanation of the pseudogap observed
in photoemission.
We also discuss the relation of the pseudogap observed in the spectral
function to the pseudogaps observed in the magnetic susceptibility.
\end{abstract}

\begin{multicols}{2}

\section{Introduction}

In this note we report results of a theoretical study of a ``spin liquid''
with a van Hove singularity near the Fermi surface.
By the term ``spin liquid'' we mean a liquid of charge $0$ spin $1/2$ fermions
filling a large Fermi sea and coupled by a singular gauge field
interaction \cite{Baskaran87,Ioffe89} and by additional non-singular
interactions.
We determine exactly the effect of the van Hove singularity on the fermion
gauge-field physics and treat the additional interactions by a "leading
logarithm" renormalization group analysis.

The problem of a spin liquid with a van Hove singularity is of interest on
experimental and theoretical grounds.
The experimental motivation for describing the high $T_c$ superconductors as
spin liquids has been discussed at length elsewhere\cite{Lee90}.
The importance of van Hove singularities has been dramatically confirmed by
recent angle resolved photoemission measurements of the doping dependence of
the Fermi surface\cite{Dessau95}.
The qualitative doping dependence expected theoretically for non-interacting
electrons  is sketched in Fig. 1.
For overdoped samples (dashed curve) the Fermi surface is closed and
electron-like.
In a non-interacting model, the Fermi surface would grow as electrons are
added, until it reached the van-Hove points.
A Fermi surface passing through the van Hove points is shown as the
solid line in Fig. 1.
Instead, the experimental result\cite{Dessau95} is that as more electrons are
added, the material develops a gap near the van Hove points.
States in the vicinity of the van-Hove points are pushed away from the Fermi
surface.
For ``underdoped'' materials no states with energies near the chemical
potential are observed near the zone edges.
States are observed near the chemical potential only in the disconnected
regions along the zone-diagonal shown as solid arcs in Fig. 1.
The existence and consequences of this ``non-Luttinger'' Fermi
surface require theoretical explanation.

The subject of non-Luttinger Fermi surfaces has attracted substantial
theoretical attention.
The general approach has been to start with fermions with a large (Luttinger)
Fermi surface and then to invoke a physical mechanism to open a ``pseudogap''
which eliminates part or all of the Fermi surface.
Three classes of mechanisms have been extensively considered:
(i) quasi-long-ranged antiferromagnetic spin fluctuations,
(ii) the ``d-wave RVB'' state, and (iii) the ``staggered flux phase''.
None has proven completely satisfactory; we discuss each in turn.

The logic behind the antiferromagnetic spin fluctuations approach
is that static antiferromagnetic order at wavevector $\bf{Q}$ leads to Bragg
scattering at $\bf{Q}$ which may open a gap over all or part of the Fermi
surface.
Schrieffer and co-workers have argued that sufficiently slowly varying
antiferromagnetic fluctuations with a sufficiently long correlation length may
also lead, if not to a gap, at least to a rather strong suppression of the
Fermi surface density of states\cite{Schreiffer88}.
A difficulty with this picture is that magnetic instabilities of a wide
variety of models have been investigated; pseudogaps have only been found
in parameter regimes leading to long-ranged order at $T=0$
\cite{Lee73,Sadovskii74}.
The essential reason is that quasistatic (i.e. frequency much less than
temperature) spin fluctuations are required for pseudogap
formation \cite{Schreiffer88,Lee73,Sadovskii74}.
In metallic and superconducting high $T_c$ materials, the spin fluctuations
observed by NMR have characteristic frequency scale of order $T$ or
greater \cite{NMR} (corresponding to no long range order at $T=0$) and are not
sufficient to open a pseudogap in the models which have been considered.

\vspace{-.5in}
\centerline{\epsfxsize=8cm \epsfbox{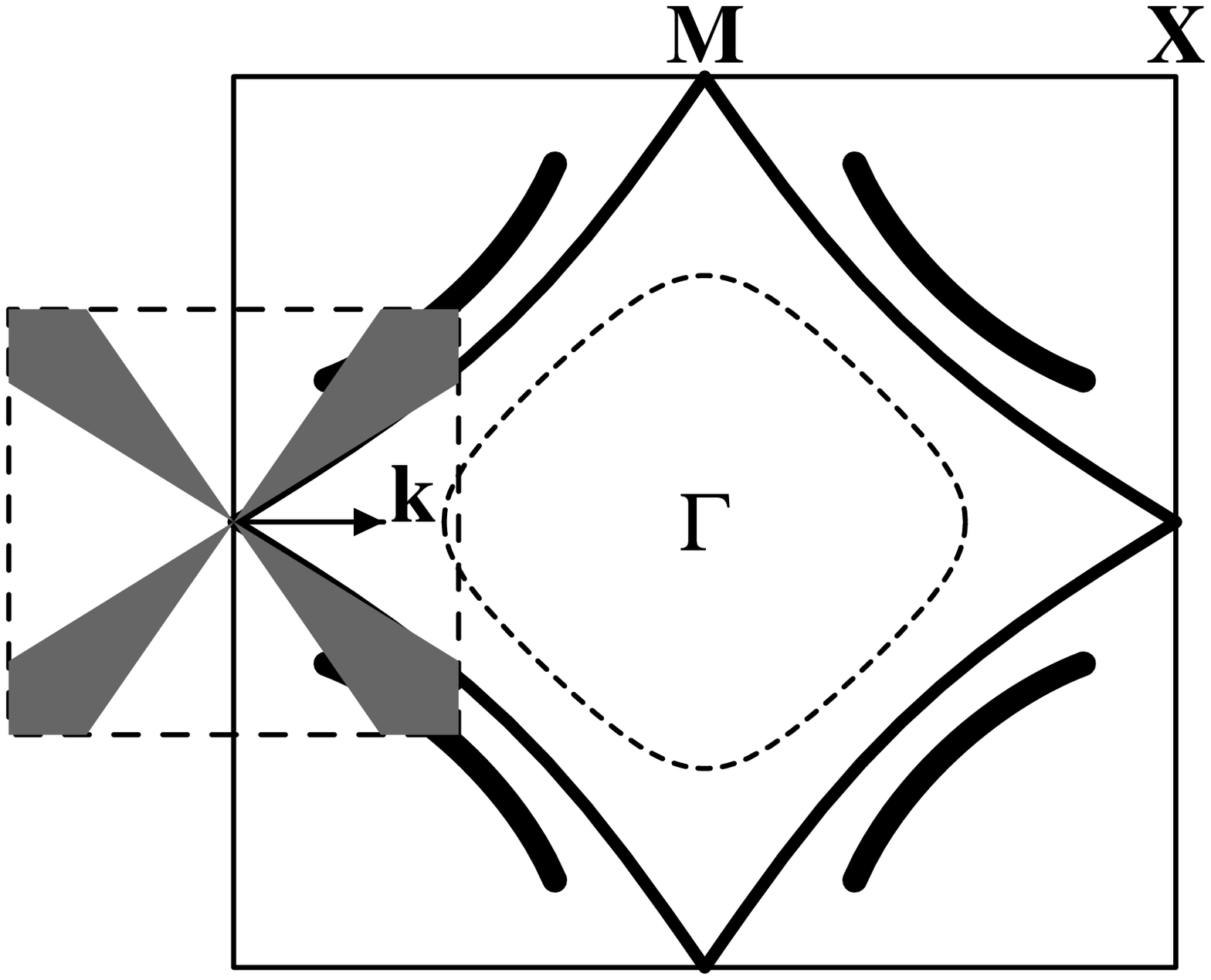}}
\vspace{-.75in}
{ \small
Fig 1.
Large square: Brillouin zone for fermions.
Solid line: Fermi surface passing through the van Hove point.
Dashed line: Fermi surface of overdoped material.
Heavy arcs: region where Fermi surface was observed in photoemission
experiment in underdoped $BSCCO$.
Dashed square: inset showing  phase space for gauge field fluctuations.
Gauge field fluctuations with momentum $\bf k$ in one of the four shaded
regions are described by the conventional overdamped propagator; fluctuations
with $\bf k$ outside of these regions are much larger because there is no part
of the Fermi surface tangent to $\bf k$.
}
\vspace{.1in}

An alternative model for the origin of the pseudogap is the ``d-wave
RVB'' model.
This is a mathematical expression of Anderson's original insight
\cite{Anderson87} that in models involving a strong on-site repulsion and a
density near one electron per site, singlet pairing and antiferromagnetism do
not differ much in their short-ranged correlations and energies.
To implement Anderson's idea one assumes that ``spin-charge separation''
occurs, and that the spin degrees of freedom are described by charge $0$
fermionic ``spinons''.
These fill a Fermi sea with a large (Luttinger) Fermi surface.
This Fermi sea of spinons may be derived in a mean field theory, ``the uniform
RVB phase'',\cite{Ioffe89,Affleck88} of the t-J model.
The t-J model is believed by many \cite{Lee90} but not all \cite{Varma88}
authors to contain the essential physics of high-$T_c$ superconductors, and
the uniform RVB phase is believed to be the most physically appropriate
starting point, at least for materials near the optimal doping.
To understand pseudogap formation one then considers instabilities of the
uniform RVB state, which are due to  residual interactions neglected in
the mean field theory.
Before now calculations have been based on ``generic'' Fermi surfaces without
van Hove points and have considered instabilities to antiferromagnetism, to
d-wave pairing, and to a staggered flux phase.

The antiferromagnetic instability has been considered and found not to lead to
a pseudogap for essentially the same reason as in the Fermi liquid case
\cite{Altshuler95a,Altshuler95b}.

The staggered flux phase involves the appearance of circulating spin currents.
This entails spontaneous breaking of time reversal symmetry which has been
observed  not to occur in cuprates \cite{Spielman90}.
For this phenomenological reason the staggered flux phase has been
discarded.

The d-wave RVB state naturally leads to the formation of a pseudogap, and the
resulting phenomenology provides an attractive scenario for the
cuprates\cite{Tanamoto92}.
The d-wave RVB state may be viewed as arising from a pairing instability of
the uniform RVB state; the resulting theory is very similar to that arising
from conventional superconducting pairing.
One important difference is that because the pairing involves chargeless
``spinons'' it does not lead to superfluidity or indeed any other observable
which could serve as a order parameter.
For this reason fluctuation corrections convert the pairing transition to a
smooth crossover.

A difficulty with this scenario has been pointed out by Ubbens and Lee
\cite{Ubbens94}.
Their results, we believe, are most simply interpreted as saying that the
spinon-gauge-field interaction produces a very short inelastic lifetime for
the spinons.
This inelastic scattering is so strongly pairbreaking that it completely
suppresses the d-wave pairing  instability.
Of course, a first order transition to a paired state would be possible
\cite{Ubbens94}, but is not observed.

Very recently, a model with an $SU(2)\times SU(2)$ symmetry has been
considered \cite{Wen95}.
This symmetry is in principle broken down to $SU(2)$ only at any non-zero
doping, but the symmetry breaking is argued not to be important.
In this model a staggered flux instability occurs which does not break time
reversal symmetry and can be transformed using an operation in the $SU(2)\times
SU(2)$ group to a $d$-wave pairing state.
The effects of gauge fluctuations on this state have not yet been determined.
On the mean field level it leads to a phenomenology very similar to the one we
shall derive in this paper.

All of the previously discussed theoretical calculations were based on
``generic'' models which did not contain van Hove sigularities.
In this communication we show that when the effects of proximity to van Hove
singularities are properly included, the theoretical predictions are very
substantially altered.
Most notably, near Van Hove points the  inelastic scattering is suppressed,
permitting the formation of a d-wave paired state with a gap which is non-zero
only near van Hove points.
This d-wave state should be reconsidered as a possible explanation for the
pseudogap.

Near a van Hove singularity the fermion density of states diverges, so that
even arbitrarily weak interactions can produce large effects.
This line of thinking has led to a large literature devoted to analyzing the
implications of van Hove singularities in Fermi liquid models
\cite{Lee89,Dagotto95,Newns92,Dzyaloshinskii87}.
We have already mentioned the strong phenomenological evidence for regarding
high $T_c$ materials as spin liquids\cite{Lee90}.
An additional difficulty with Fermi liquid based treatments is that the
singularities due to the van Hove points are cut off by hopping in third
dimension.
The bare value of this hopping may be determined from band structure
calculations and is not small \cite{Pickett89}.
Evidently in the actual materials the interplane coupling is
renormalized down to a very small value in $BSCCO$ and underdoped
$YBa_2Cu_3O_{7-x}$ \cite{Timusk95}; this renormalization cannot
be understood in a Fermi liquid picture but follows naturally in a spin
liquid.

The remainder of this paper is organized as follows.
In Section II we formulate and solve a model of a spin liquid with a van Hove
singularity coupled to a gauge field.
In Section III we employ a leading logarithm approximation to determine the
effects of residual non-singular interactions.
Section IV is a conclusion which summarizes the approximations employed,
results obtained  and consequencies for photoemission and other physical
properties.
An Appendix gives derivations of some results used in the body of the paper.

\section{Spinon-Gauge Model}

The spin liquid state we shall start from is the ``uniform RVB'' state.
The Hamiltonian is
\begin{equation}
\begin{array}{rl}
H & = \sum_{p\sigma} \epsilon_p c_{p\sigma}^\dagger c_{p\sigma}  \\
  & + \frac{1}{2} \sum_{pk\sigma} \vec{a}_k c_{p+k/2,\sigma}^\dagger
 [\vec{v}_{p+k/2} + \vec{v}_{p-k/2}] c_{p-k/2,\sigma} \\
 & + W \sum_p c_{p_1 \sigma}^\dagger c_{p_2 \sigma} c_{p_3 \sigma}^\dagger
 c_{p_4 \sigma} \delta ( \Sigma p_i )+
 \frac{1}{4g_0^2} f_{\mu \nu}^2
 \end{array}
\label{H}
\end{equation}
This Hamiltonian is the usual one (see, e.g. \cite{Baskaran87,Altshuler94}),
however we have written the fermion-gauge-field coupling in its general form.
Here $\vec{v} (p)=\partial \epsilon_p / \partial \vec{p}$ and
$f_{\mu \nu} = \epsilon_{\mu \nu} \partial_\mu a_\nu$.
It represents the low energy spin degrees of freedom of the $t-J$ model.
Low energy in this context means energies less than $J$, which in high $T_c$
materials is about $0.15\;eV$.
The coefficient $g_0^{-2}$ contains the contributions to the gauge-field
stiffness from the higher energy spin degrees of freedom which were integrated
out in the derivation of Eq. (\ref{H}).
It may be expressed in terms of the high energy part of a six spin correlation
function and  is  of the order of $J$ in magnitude; it is discussed further in
Appendix A.

We assume the fermion spectrum is
\begin{equation}
\begin{array}{rl}
\epsilon_p = & - 2t [cos (p_xa) + cos(p_ya) ] \\
	& - 4t' \cos(p_xa) \cos(p_ya)  -\mu - 4t'
\end{array}
\label{epsilon_p}
\end{equation}
Here $a$ is the lattice constant, $\mu+4t'$ is the chemical potential with
$\mu=0$ corresponding to the van Hove point, $t$ is a first neighbor hopping
due mostly to the superexchange $J$ with an additional contribution coming
from a bandstructure $t_{band}$ renormalized by the hole density and the
parameter $t^{\prime}$ is derived from further neighbor hopping in the
underlying bandstructure also renormalized by the hole density; roughly, we
expect
\begin{eqnarray}
t = \Theta J + t_{band} \delta
\label{t} \\
t' = t_{band}' \delta.
\label{t'}
\end{eqnarray}
Here $\Theta$ is a number of the order of unity, in the large $N$ limit
$\Theta=2/\pi^2$.
It is believed that $t' < 0$ in high $T_c$ superconductors \cite{Pickett89}.
Note that there is no coherent hopping of spinons in the third dimension;
in the model such hopping cannot occur unless the band-structure $t_\perp
\gtrsim t_{band}$.
The derivation of the spinon $\epsilon_p$ from the $t-J$ model is outlined in
Appendix A.

Near a van Hove point (e.g. $p_x = \pi , \; p_y =0)$ we have
\begin{equation}
\epsilon_p = -u_0 (p_x^2 - \alpha^2 p_y^2 )
\end{equation}
with
$u_0= (t+ 2t') a^2$ and
$\alpha^2 = \frac{t-2t'}{t+2t'}$.
If $t' \neq 0$ then $\alpha^2 \neq 1$ and the energy contour which passes
through the van Hove points is not nested.
An example of such a contour is shown as the solid line in Fig. 1.
At the van Hove points the velocity vanishes, implying a diverging density of
states and a vanishing of the fermion-gauge-field coupling.

We now consider the fermion-gauge field interaction.
Previous work\cite{Altshuler94} has shown that it is correct to analyze this
interaction in two steps using a loop expansion controlled by the parameter
$N$, the fermion spin degeneracy.
In the physical problem $N=2$.
It has been shown that results obtained at leading order in a $1/N$ expansion
are not significantly changed at higher orders.
First, one constructs the renormalized gauge-field propagator D by
dressing the term $f^2_{\mu \nu}/g_0^2$ by the fermion transverse
current-current polarizibility $\Pi = \int v v G G$.
Second, one uses this to compute the fermion self-energy.
For a closed Fermi surface far from van Hove points,
$\pi ( \omega, \vec{k} ) = \frac{p_0| \omega |}{\pi |k|} + \chi k^2$.
For a given direction of $\vec{k}$, the dissipative term $| \omega | / |k|$
comes from fermions near the points on the Fermi surface which are tangent to
$\vec{k}$; $p_0$ is the curvature of the Fermi surface at these points and
$\chi$ is the diamagnetic susceptibility of the fermions.

The presence of van Hove singularities leads to two effects.
First, as can be seen from Fig. 1, there is a range of directions of $\vec{k}$
which are not tangent to any point on the Fermi surface.
For these directions, the dissipative term in $\Pi(\omega,k)$, $\Pi_{diss}$,
is much smaller.
To display the behavior of the dissipative term it is convenient to define
coordinates $k_{\pm} = k_x \pm \alpha k_y$.

If $\frac{1-\alpha}{1+\alpha} < |k_+/k_-| < \frac{1+\alpha}{1-\alpha}$, then
\begin{equation}
\Pi_{diss} ( \omega , k) =
\frac{\alpha N | \omega |}{4 \pi}\left( \frac{k^2}{k_x^2-\alpha^2k_y^2}
+\frac{k^2}{k_y^2-\alpha^2k_x^2} \right)
\end{equation}
while if $|k_+ /k_- |$ is outside this range then
\begin{equation}
\Pi_{diss} (\omega ,k)= \frac{N p_0 |\omega |}{2 \pi |k|}
\end{equation}
Second, states in the vicinity of the van Hove point produce a
negative, divergent contribution to $\chi$, so
\begin{equation}
\chi = - \frac{N u_0}{6 \pi^2} \; \ln \; \epsilon_F / \Lambda + \chi_{reg}
\label{chi}
\end{equation}
Here the logarithm cutoff $\Lambda={\mbox max}(T,\mu,u_0 k^2)$ and
$\chi_{reg}$ is the contribution from fermions far from the van Hove
points.
The total $\chi$ calculated for non-interacting fermions
with $N=2$ moving in the band
structure defined by Eq. (\ref{epsilon_p}) is shown in Fig. 2; as seen in this
Figure $\chi$ near half filling is negative and of small magnitude with a weak
divergence near van Hove points as expected from Eq. (\ref{chi}).
The full gauge propagator is thus
\begin{equation}
D(\omega,k) = \frac{1}{\Pi_{diss} + (g_0^{-2} + \chi)k^2}
\end{equation}
The divergence of $\chi$ implies that in the vicinity of the van Hove
singularity the uniform RVB state is unstable to a state of non-zero flux, at
a temperature
\begin{equation}
T_{flux} \sim \epsilon_F \; \exp -
\left [ \frac{6 \pi^2(\chi_{reg}+g_0^{-2})}{N u_0} \right]
\label{T_flux}
\end{equation}
unless a different instability occurs first.
States of non-zero flux break time reversal invariance; as there is no
evidence that this occurs in high $T_c$ materials we shall assume that
$\chi_{reg}+g_0^{-2}$ is sufficiently large  that $T_{flux}$ is negligibly
small.
This assumption is consistent with the previously mentioned theoretical
estimate $g_0^{-2} \sim J $ which is greater than the typical values of $\chi
\sim 0.1 t \sim 0.1 J$ shown in Fig. 2.
Also, a previous analysis of the $T$-dependence of the resistivity at $T>100\;
K$ predicted by Eq. (\ref{H}) found $g_0^{-2}+\chi_{reg} \approx 500 a^2 \; K
\sim Ja^2/3$.
These estimates combined with the large numerical factor in Eq. (\ref{T_flux})
imply that the instability will occur only very close to the van Hove point,
and only at very low temperature.

At scales of interest we may then neglect the logarithm and write
\begin{equation}
D(\omega,k) = \frac{1}{\Pi_{diss} + k^2 /g^2}.
\end{equation}
where $g^{-2}=g_0^{-2} + \chi \approx 500 a^2\; K$.
We now use this $D(\omega,k)$ to calculate the fermion self energy.
For fermions far from the van Hove points the calculation is identical to that
given in previous work\cite{Altshuler94}.
The kinematics of a scattering event imply that the fermion is scattered
parallel to the Fermi surface.
For these momenta one must use the $|\omega| /|k|$ form of $\Pi_{diss}$,
leading to
\begin{equation}
\Sigma_{far} (\epsilon , \vec{p}) = [ \omega_0 ( \vec{p})]^{1/3} \epsilon^{2/3}
\label{Sigma_far}
\end{equation}
Here
\begin{equation}
\omega_0(\vec{p})=\frac{v_F^3(\vec{p})g^4}{\pi^2 p_0(2 \sqrt3 )^3}
\label{omega_0}
\end{equation}
at leading order in $N$.
The Fermi velocity $v_F(p)$ vanishes linearly as one approaches the van Hove
point, implying $\Sigma_{far}$ does also.
Note, also that $1/p_0$ vanishes for a nested Fermi surface; from Eq.
(\ref{t'}) we conclude that $1/p_0 \propto \delta$, so $\omega_0 \propto
\delta$.

\vspace{-.5in}
\centerline{\epsfxsize=8cm \epsfbox{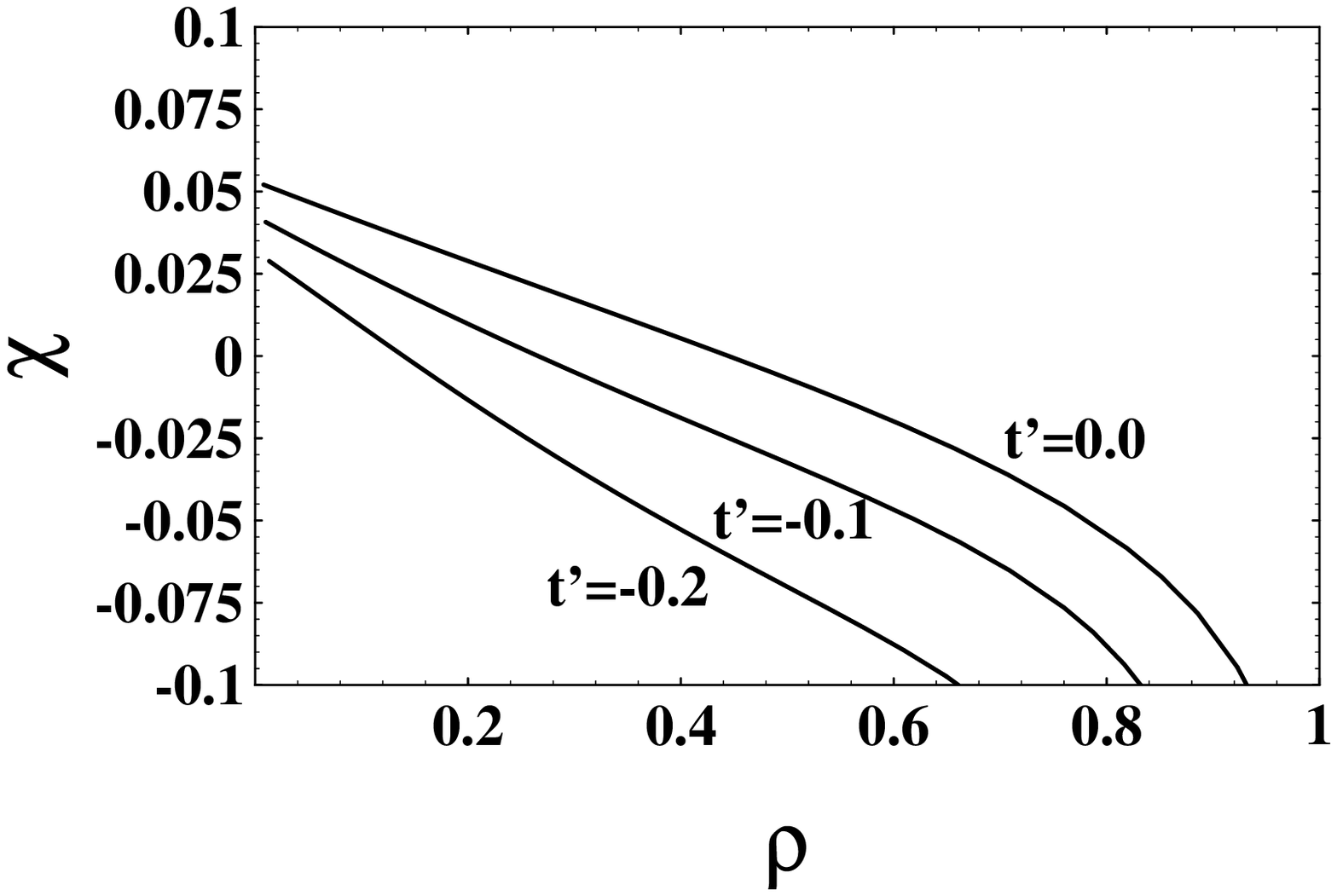}}
\vspace{-.75in}
{\small
Fig 2.
Diamagnetic susceptibility of non-interacting Fermi gas with spectrum
given by Eq. (2) as a function of filling, $\rho$, for $t=1$ and several
values of the next neighbour hopping, $t'$.}
\vspace{.1in}

For fermions near the van Hove point, scattering processes involving the other
form of $\Pi_{diss}$ become allowed.
These lead to a contribution
\begin{equation}
\Sigma_{near} (\epsilon,p)= -\frac{1}{2\pi^2}
	\!\left(\! \ln\frac{\epsilon_F}{\epsilon} \ln\frac{u}{\pi \chi}
	\!\right)\!
	\left[\! i A \epsilon \!-\! Bu(p_x^2\! - \! \alpha^2 p_y^2) \! \right]
\label{Sigma_near}
\end{equation}
where the numerical coefficients $A=\ln\frac{1+\alpha}{1-\alpha}-1$,
$B=3\ln\frac{1+\alpha}{1-\alpha}-1$.
$\Sigma_{near}(\epsilon,p)$ becomes important for $\epsilon<\epsilon^*$ where
$\frac{1}{2\pi^2} \ln\frac{\epsilon_F}{\epsilon^*} \ln\frac{u}{\pi \chi} = 1$;
$\ln\frac{u}{\pi \chi} \approx 1$, so $\ln\frac{\epsilon_F}{\epsilon^*}$ is
within a factor of two of $\ln\frac{\epsilon_F}{T_{flux}}$, thus, for
consistency, we must assume $\Sigma_{near}(\epsilon,p)$ is negligible.
The instability in $\Sigma_{near}(\epsilon,p)$ reinforces that in $\chi$.
To study this instability more carefully we have written and solved coupled RG
equations for u and g;
we find that the more careful considerations do not change the estimate
of the scale $T_{flux}$ where these effects become important.
Thus we assume
\begin{equation}
\Sigma(\epsilon ,\vec{p}) \cong \Sigma_{far} (\epsilon ,\vec{p});
\label{Sigma}
\end{equation}
in other words the self energy due to fermion-gauge field  scattering is
important far from the van Hove points and unimportant near them, provided
that the instability towards the flux phase may be neglected.

\section{Effect of Non-Singular Interactions}

We now turn to the residual, non-singular interactions.
It has already been shown that the short range interaction between fermions
far from van Hove points is renormalized to zero by the gauge
field \cite{Altshuler94,Polchinski94}.
We thus need only to consider processes involving fermions near van Hove
points.
For these processes, vertex renormalizations due to the gauge field lead to
the same logarithms we have already agreed to neglect.
The theory we must consider is therefore of fermions near the van Hove points,
with self energy $\Sigma_{far}(\epsilon,p)$, coupled by short range
interactions which may be parametrized by four constants: particle-particle and
particle-hole interactions with momentum transfer near zero or $G =(\pi ,\pi)$.

The theory has some formal similarity to theories of fermions in one dimension.
In one dimensional models one considers only fermions near the Fermi surface
of the left and right hand branches of the dispersion relation; here one
considers only fermions near van Hove points.
In one dimensional models a weak coupling leading logarithm approximation
exists because some particle-hole and particle-particle propagators
diverge logarithmically due to kinematics in one dimension.
In the present model the divergent density of states at the van Hove
singularity similarly leads to  logarithmic divergences of susceptibilities.

The theory is more complicated than theories of fermions in one dimension
because there are two small scales: $\omega_\alpha=\frac{1-\alpha}{1+\alpha}
\epsilon_F$ which will be
shown below to set the scale at which deviations from the perfect nesting
become important, and $\omega_0$ which sets the scale at which gauge field
effects become important.
Both $\omega_\alpha$ and $\omega_0$ are proportional to $\delta$ as
previously discussed; because they  have the same doping dependence and
$\omega_0$ is small even at large doping due to the large value of the gauge
stiffness $g^{-2}$ in Eq. (\ref{omega_0}), we believe that $\omega_0 <
\omega_\alpha$ is the only relevant case.

The charges in the antiferromagnetic and $d$-wave pairing channels diverge.
At scales larger than $\omega_\alpha$ and $\omega_0$ both channels
have a $\ln^2(1/\epsilon)$ divergence.
At scales less than $\omega_\alpha$ the antiferromagnetic divergence
becomes $\ln(1/\epsilon)$.
At scales less than $\omega_0$ the coefficient of the $\ln^2(1/\epsilon)$ in
the superconductivity channel (and, if $\omega_0>\omega_\alpha$, in
the antiferromagnetic channel) is reduced, because the gauge field produces a
strong inelastic scattering in some regions of momentum space.

In the remainder of this Section we present the results of the leading
logarithm calculations.
We assume throughout that at most one coupling becomes large.
Especially in the regime of  $\ln^2(1/\epsilon)$ renormalizations the problem
of coupled charges is very involved and has been treated elsewhere
\cite{Dzyaloshinskii87}.

\subsection{Antiferromagnetic instability}

We proceed to construct leading logarithm RG equations.
There is no logarithm in the small momentum transfer particle-hole
susceptibility, so the corresponding charge is not renormalized.
Moreover, in contrast to 1D, the small momentum transfer particle hole
susceptibility has only a weak singularity as $\omega=vq$, so these processes
can be neglected.
The $\bf G$ momentum transfer particle-hole processes lead to logarithmic
divergences.
For a nested Fermi surface ($\alpha^2=1$) the two momentum integrals are {\em
each} logarithmically divergent leading to a $\ln^2(1/\epsilon)$
renormalization of the charge.
For a non-nested Fermi surface ($\alpha^2 \neq 1$) one logarithm is cut off by
$(1-\alpha^2)$ if it has not already been cut off by temperature.
The crossover occurs at the scale $\omega \approx \omega_\alpha$.
The existence of a logarithmic divergence in the particle hole bubble means
that the charge, $g_{AF}$, associated with $\bf G$ momentum  transfer
particle-hole processes grows.
In the leading logarithm approximation we find that antiferromagnetic charge
is renormalized by
\begin{equation}
\frac{\delta g_{AF}}{g_{AF}} \!=\! \left\{
\begin{array}{lcl}
\frac{g_{AF}}{8\pi^2 u} \ln^2 \frac{\epsilon_F}{T}
&\;\;& T >\omega_0,\omega_\alpha \\
\frac{g_{AF}}{8 \pi^2u} \ln\frac{1+\alpha}{1-\alpha}
	\ln\frac{\epsilon_F \omega_\alpha}{T^2}
&\;\; &T < \omega_\alpha
\end{array}
\right.
\label{g_AF}
\end{equation}
The second of these formulas was derived on the
assumption that $\omega_\alpha > \omega_0$;
if not an additional crossover occurs.
The formulas relevant to this case will be presented and discussed in the next
subsection treating pairing.
The renormalization of $g_{AF}$ calculated from Eq. (\ref{g_AF}) is shown as
the dashed line in Fig. 3.

\vspace{-.5in}
\centerline{\epsfxsize=8cm \epsfbox{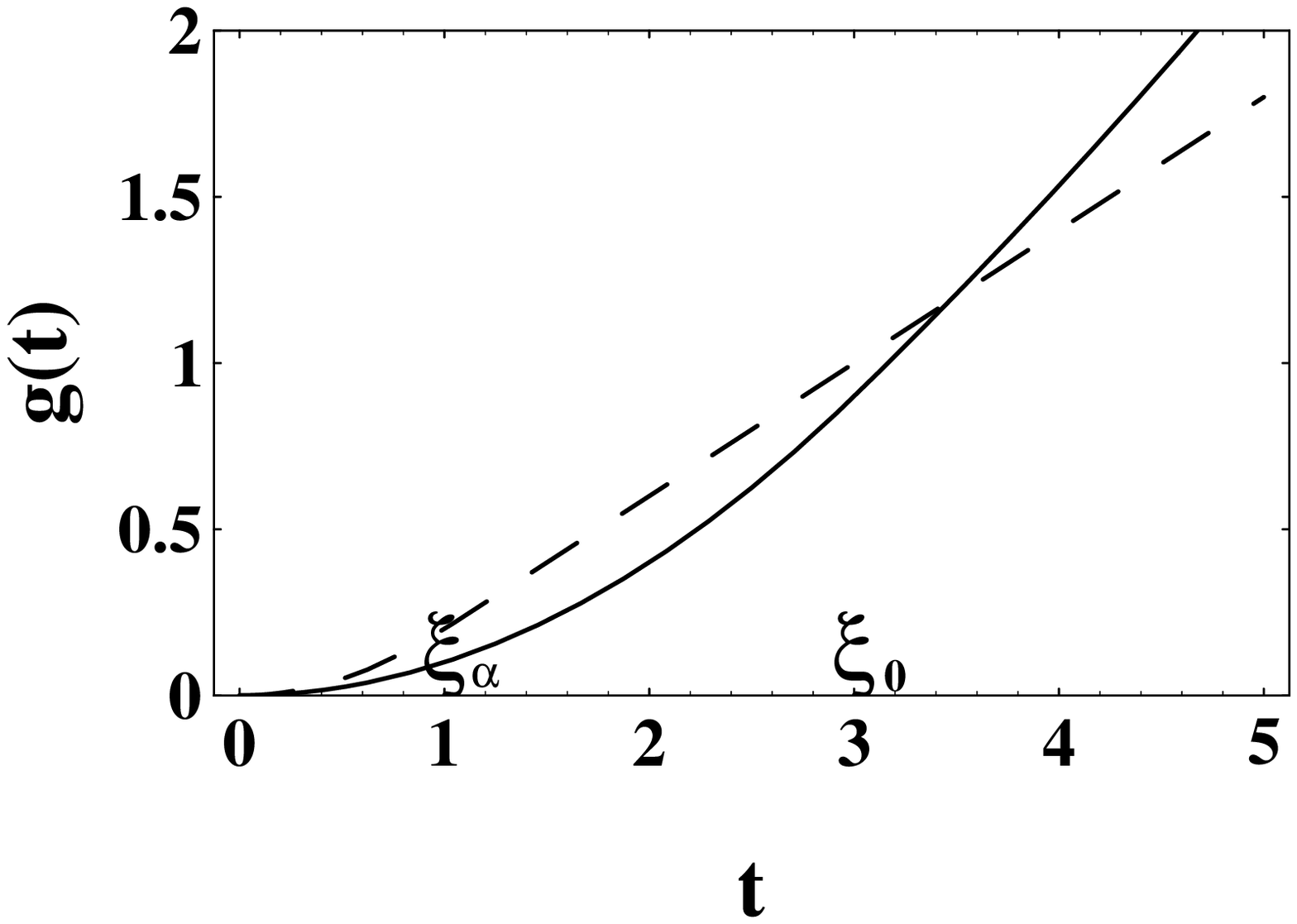}}
\vspace{-.75in}
{\small
Fig 3.
Renormalization of the interaction constants $g=\frac{\delta g_{AF}}{g_{AF}}$,
$\frac{\delta g_{p}}{g_{p}}$ plotted against $t=\ln(\epsilon_F/T)$.
Solid line: $\frac{\delta g_{p}}{g_{p}}$ calculated from Eq. (\ref{C(T)}) with
initial condition  $\frac{g_p}{8\pi^2 u}=0.1$.
Dashed line: $\frac{\delta g_{AF}}{g_{AF}}$ calculated from Eq. (\ref{g_AF})
with initial condition $\frac{g_{AF}}{8\pi^2 u}=0.2$.
}
\vspace{.2in}

We define $T_{AF}$ as the scale at which $\frac{\delta g_{AF}}{g_{AF}}$ becomes
of the order of unity.
At scales larger than $T_{AF}$, the effects of $g_{AF}$
are negligible within the weak coupling approximation.
At scales of order $T_{AF}$, the antiferromagnetic susceptibility
becomes large, the renormalization of the fermion propagator
becomes large, and perturbation theory breaks down.
It is difficult to make definite predictions for temperatures lower than
$T_{AF}$ because one is then dealing with a strongly interacting model.
We can imagine two scenarios - a crossover to the d-wave RVB regime to be
discussed below or a crossover to an antiferromagnetic regime involving
critical fluctuations.
We discuss the antiferromagnetic regime here, focussing on whether the
important fluctuations are quantum or classical and whether they can open a
gap in the fermion spectrum.

It is helpful to compare the present calculation to the well known BCS theory
of superconductivity which also has a logarithmic divergence of the coupling
$g_{sc}$.
In superconductivity the important fluctuations are classical (i.e.
involve modes with energies less than $k_B T$) and are generally weak but grow
as $T$ approaches $T_c$.
These fluctuations are described by the classical Landau theory
\begin{equation}
F_{sc} = \nu \int d^dk (\tau+\xi_0^2 k^2) \Delta^2_k + \beta_{sc} \int d^d r
	\Delta^4_r
\label{F_sc}
\end{equation}
where $\nu \sim 1/\epsilon_F$ is the density of states,
the coherence length $\xi_0=v_F/T_c$ and
$\beta_{sc} \sim \nu/T_c^2$.
The renormalization of $\beta_{sc}$ may be estimated to be
\cite{Landau-Lifshitz}
\begin{equation}
\frac{\delta \beta_{sc}}{\beta_{sc}} \sim
	\left( \frac{T_c}{\epsilon_F} \right)^{d-1} \tau^\frac{d-4}{2}
\label{delta_beta_sc}
\end{equation}
where $\tau=\frac{T-T_c}{T_c}$.
This estimate shows that at the scale $\tau \approx 1$ where the coupling
$g_{sc}$ has scaled to the order of unity, thermal fluctuations are negligible
in dimension $d>1$.
By continuity quantum fluctuations must be also negligible at this scale.
This may be seen directly: at scales larger than $T_c$, the $\phi^4$ vertex
of the quantum Landau theory
$\beta_{sc}(\epsilon,k) \sim \frac{1}{\epsilon^2 + (vk)^2}$ and the Cooper
propagator has negligible momentum dependence so the leading correction
$\delta \beta_{sc}/\beta_{sc} \sim (T_c/\epsilon_F)^{d-1}$.
Thus we see from the quantum calculation that $d=1$ is the marginal
dimensionality for the quantum fluctuations; this fact is revealed in the
classical calculation because in $d=1$ the Ginzburg parameter
$(T_c/\epsilon_F)^{d-1}$ ceases to be small.

In the antiferromagnetic problem of interest here, the marginal dimension is
$d=2$ because the fermion spectrum is $\epsilon \propto p^2$ rather than
$\epsilon \propto p$.
The free energy for the classical fluctuations is
\begin{equation}
F_{cl} = \frac{\ln \frac{1+\alpha}{1-\alpha}}{u} \int (d^2k) \phi_k^2
	\left[ F_1 \tau + F_2 \frac{uk^2}{T_{AF}} \right]
	+ \beta_{AF} \int \phi^4_r d^2 r
\label{F_cl}
\end{equation}
with $\beta_{AF} = F_3 /(uT_{AF}^2)$ and $\tau=(T-T_{AF})/T_{AF}$ ($F_i$ are
numerical coefficients of the order of unity). From the
free energy (\ref{F_cl}) one finds
\begin{equation}
\frac{\delta \beta_{AF}}{\beta_{AF}} \sim \frac{G_i}{|\tau|}, \;\;
	G_i = \ln^{-2} \frac{1+ \alpha}{1-\alpha}
\label{delta_beta_AF}
\end{equation}
We see that $d=2$ is the critical dimension because the only parameter (apart
from $\tau$) controlling the fluctuation correction is
$\ln\frac{1+\alpha}{1-\alpha}$ which does not depend on $T_c/\epsilon_F$.
The calculation of the leading quantum fluctuation corrections shows that these
also are small only by a power of $\ln\frac{1+\alpha}{1-\alpha}$.
We cannot proceed further in the general case, but in the limit $\alpha
\rightarrow 1$ we can use the $\ln\frac{1+\alpha}{1-\alpha}$ to control the
calculation assuming that the antiferromagnetic charge is the only relevant
one.
However, a theory of the general $\alpha \rightarrow 1$ case found more
than one relevant charge and the resulting theory of the competing
instabilities is very involved \cite{Dzyaloshinskii87}.

In our $\alpha \rightarrow 1$ limit, the only important fluctuations are
classical.
There is no long range order at any $T>0$; rather, for $T<T_{AF}$,
$\Delta$ crosses over to $\Delta \sim \exp\left(\frac{G_i}{\tau}\right)$
implying an exponentially growing correlation length $\xi_{AF} \sim
\Delta^{-1/2}$.
These fluctuations are thus quasistatic and long ranged and in particular have
energy much less than the typical fermion energy $k_B T$ and momentum much
less than the typical fermion momentum, $p \sim \sqrt{T/u}$, so as far as the
fermions are concerned these fluctuations may be treated as static periodic
scatterers, and  lead to a gap $\Delta_F \approx T_{AF} \sqrt{-\tau}$.
The magnetically induced fermion pseudogap found in this model is not relevant
to the underdoped high $T_c$ materials because the fluctuations producing the
pseudogap would also lead to a very rapid $T$-dependence of the $Cu$ NMR
relaxation rates proportional to powers of $\xi_{AF}\sim \exp(G_i T_{AF}/T)$;
such a rapid temperature dependence is not consistent with $Cu$ relaxation
rate measurements in high $T_c$ materials \cite{NMR}.

\subsection{Superconducting instability}

We now consider particle-particle processes.
As usual the leading divergence happens in the $q=0$ momentum channel.
The Cooper propagator, $C(T)$, is
\begin{equation}
C(T) = \int \frac{d^2pd\epsilon}{(2\pi)^3}
\frac{g_p}{\epsilon_p^2+(\epsilon +\omega_0(p)^{1/3}\epsilon^{2/3})^2}
\label{C(T)_0}
\end{equation}
Near the corners $\omega_0(p_{\pm}) \sim (p_+p_-)^3$
and $\epsilon_p = - u p_{+}p_{-}$, so one finds
\begin{equation}
C(T) \! \approx \! \left\{ \begin{array}{ll}
\frac{g_p}{8\pi^2 u}\ln^2 \frac{\epsilon_F}{T}\hspace{1in} & T \geq \omega_0 \\
\frac{g_p}{24 \pi^2 u} \! \left(\ln^2 \!\frac{\epsilon_F^3}{\omega_0^2T} \!-\!
	6\! \ln^2\!\frac{\epsilon_F}{\omega_0} \right)
		& T < \omega_0
\end{array} \right.
\label{C(T)}
\end{equation}
Here $g_p$ is the pairing coupling constant.

The $\ln^2\frac{\epsilon_F}{T}$ divergence of $C(T)$ is due to the divergence
of the density of states and the vanishing of the inelastic lifetime
near the corners.
The change in $C(T)$ as $T$ is reduced below $\omega_0$ reflects the
pairbreaking effect of the gauge fluctuations in the regions away from van
Hove points.
One sees immediately from (\ref{C(T)_0}) that $C(T)$ is non-divergent in models
where $\omega_0 (p) \sim const$, as shown by Ubbens and Lee\cite{Ubbens94}.
Ubbens and Lee derived their results by an energy argument which demonstrated
that a second order pairing transition is not possible in models with
$\omega_0(p) \sim const$.
We believe the argument using the gap equation is equivalent but more
transparent, however we note that if we apply their energy arguments to the
present model we discover that a second order transition is possible.

The $\ln^2\frac{\epsilon_F}{T}$ divergence of $C(T)$ means that pairing is
the instability which dominates the logarithmic antiferromagnetic and flux
instabilities in weak coupling.
However, because the interactions are repulsive the pairing instability occurs
in the d-wave channel.
Formally, one must consider the pairing amplitudes $\Delta_{11}$ and
$\Delta_{22}$ near the two inequivalent corners 1 and 2 and the appropriate
interaction amplitudes $g_{11}$, $g_{12}$.
The gap equation is dominated by the corners, so becomes a $2 \times 2$ matrix
equation
\begin{equation}
\left( \begin{array}{c} \Delta_{11} \\ \Delta_{22} \end{array} \right)
	= - C(T)
\left ( \begin{array}{cc}
g_{11} & g_{12} \\
g_{12} & g_{11} \end{array} \right)
\left( \begin{array}{r} \Delta_{11} \\ \Delta_{22} \end{array} \right)
\end{equation}

The growing eigenvalue is for $\Delta_{11} = - \Delta_{22}$ (corresponding to
d-wave symmetry) and requires that $g_{p} \equiv g_{12}-g_{11}>0$.
The pairing scale $T_p$ at which the interaction becomes of order unity is
given by
\begin{equation}
T_p  \! \approx \! \left\{ \begin{array}{ll}
\epsilon_F \exp -\sqrt{ \frac{8\pi^2u}{g_{p}}}
	& T_p > \omega_0 \\
\frac{\epsilon_F^3}{\omega_0^2} \exp -\sqrt{ \frac{24\pi^2u}{g_{p}}+
	6 \! \ln^2\frac{\epsilon_F}{\omega_0} }
	& T_p < \omega_0
\end{array} \right.
\label{T_p}
\end{equation}

In a model with point-like interactions, $g_{12}=g_{11}$; it is then necessary
to go to higher order.
Contributions to the pairing instability of order
$g_{12}^2\ln^3(\epsilon_F/T)$ exist and, for sufficiently small coupling
dominate over the AF instability.
The renormalization of $g_p$ is shown in Fig. 3.

By repeating the arguments of the previous Section we see that the Ginsburg
parameter is $G_i \approx 1/ln (\epsilon_F/T_p)$, so
mean field theory provides a reasonable description at temperatures less than
$T_p$ in weak coupling.
There is, however, one important caveat:
mean field theory predicts a phase transition to a paired state $T_p$; in
reality this is not a true transition.
The interaction between vortices of the pairing field is not logarithmic
because of screening by the gauge field, so as long as spin-charge separation
occurs vortices exist and prevent long range order.
However, the small value of the Ginzburg parameter implies that the number of
vortices present is very small, so the magnitude of the pairing field and the
implications of the pairing for the physical properties are well described by
the mean field theory.
In particular, in regions of momentum space where the pairing amplitude is
appreciable there will be a strong suppression of the electron spectral
function.

We now discuss the momentum space structure of the gap function focussing on
the physically relevant $T_p < \omega_0$ case.
The gap equation is
\begin{equation}
\Delta_{\epsilon,p} = \int \frac{ g_{pair}(\epsilon,p) \Delta_{\epsilon',p'}
	(dp')}
	{(\epsilon'+\omega_0^{1/3}(p') \epsilon^{2/3})^2 +
	(\epsilon_{p'} + \Delta_{\epsilon',p'})^2}
\label{Delta_eq}
\end{equation}

The integral in (\ref{Delta_eq}) is dominated by the van Hove point which
allows us to ignore the $p'$, $\epsilon'$ dependence of $g_{pair}$ in
(\ref{Delta_eq}).
For $p$ away from the van Hove points $g_p(\epsilon,p,p')$ is suppressed at
low scales by the gauge field fluctuations \cite{Altshuler94}, roughly
$g_{pair}(\epsilon,p)=g_p(\frac{\epsilon}{\omega_0(p)})^\kappa$ for
$\epsilon<\omega_0(p)$
with $\kappa > 2/3$.
Eq. (\ref{Delta_eq}) implies
\begin{equation}
\Delta_{\epsilon,p} = \left\{
\begin{array}{ll}
\Delta \hspace{0.2in} & \omega_0(p) < \Delta \\
\left( \frac{\epsilon}{\omega_0(p)} \right)^\kappa
	\Delta \hspace{0.2in} & \omega_0(p) > \Delta
\end{array}
\right.
\label{Delta}
\end{equation}
Here $\Delta \sim T_p$ is the pairing amplitude at the van Hove points.
Clearly, as one moves far enough from the van Hove points so that
$\omega_0(p) > \Delta$, the pairing amplitude drops rapidly and becomes less
than the scattering rate $\omega_0^{1/3}(p') \epsilon^{2/3}$ due to the gauge
field; these portions of the Fermi surface may be regarded as gapless.

\section{Conclusion}

We have studied theoretically a model of a spin liquid with a Fermi surface
which passes near a van Hove singularity.
We considered two sorts of interactions: the singular gauge interaction
arising from the spin-charge separation which established the spin liquid
in the first place, and residual short range interactions between spinons.
The gauge field interactions lead to two effects.
One is an instability at $T=T_{flux}$ to a ``flux phase'' in which time
reversal symmetry is spontaneously broken.
An expression for $T_{flux}$ is given in Eq. (\ref{T_flux}).
The numerical factors are such that $T_{flux}$ is negligible in the weak
coupling limit.
As there is no experimental evidence for time reversal symmetry breaking in
high $T_c$ materials, we assume that parameters in the physical model
are such that $T_{flux}$ is negligibly low.
This assumption allows us to neglect also the logarithmic
renormalization of the fermion dispersion shown in Eq. (\ref{Sigma}).
The second effect is a suppression, for spinons near the van Hove point, of
inelastic scattering due to the gauge field.
The physics is simple: the gauge field couples to the fermions via the
velocity; this vanishes at the van Hove point and the vanishing coupling
overcomes the diverging density of states.

We then turned to the non-singular interactions.
Processes involving spinons far from the van Hove points are known to be
renormalized to zero by the gauge-field, but near the van Hove points the
vanishing of the coupling implies that the renormalization is ineffective.
We therefore argued that we could specialize to a model involving fermions
near van Hove points coupled by non-singular interactions.
We treated this theory via weak coupling leading logarithm methods similar to
those used to study one-dimensional models.

We found diverging interactions in the $d$-wave pairing and antiferromagnetic
channels.
The coupling constant flows are shown in Fig. 3.
For sufficiently weak couplings, $d$-wave pairing dominates and a controlled
expansion based on the parameter $\ln (\epsilon_F/T)$ is possible.
Below the pairing temperature $T_p$ given in Eq. (\ref{T_p}) the $d$-wave
pairing leads to a gap in the fermion spectrum near the van Hove
points but leaves a finite region of gapless Fermi surface near the zone
diagonal.
The resulting fermion spectrum is similar to that observed in recent
photoemission experiments \cite{Dessau95}.
This pairing which eliminates some but not all of the Fermi surface will have
implications for other physical properties.
For example, the uniform spin susceptibility will decrease as $T$ is decreased
through the pairing scale, but will tend to a non-zero limit as $T\rightarrow
0$ because some of the Fermi surface remains ungapped.
Precisely this behavior occurs in $La_{2-x}Sr_xCu0_4$ at an $x$-dependent
pairing scale $T^*(x)$ varying from $T^*(x=.15) \approx 300\;K$ to
$T^*(x=.04) \approx 700\;K$.
Other properties also exhibit crossover at $T^*(x)$ \cite{Hwang94} and similar
behavior may occur in $YBaCu_3O_{6+x}$, although the data are less clear
\cite{Barzykin94}.
The identification of the RVB pairing scale $T_p$ with the empirical $T^*(x)$
was first suggested by Tanamoto {\it et al} \cite{Tanamoto92};
the new features of the present paper are a theoretical justification for the
existence of the gap despite strong inelastic scattering and the result that
the gap opens only over a small portion of the Fermi surface.

The present theory does not explain the {\em rapid} drop of $\chi_{s}(T)$ and
NMR relaxation rates observed in bilayer and trilayer materials below a spin
gap temperature $T_{SG} \sim 200K$.
Explaining these observations requires opening a gap over the whole Fermi
surface; one possibile mechanism has been discussed elsewhere
\cite{Altshuler95c}.
Finally, we note that in this theory the $d$-wave RVB state, and therefore the
fermion pseudogap, appears at a doping $\delta_g > \delta_{VH}$.
Experimentally the gap appears first for $\delta_g \sim \delta_{VH}$ and it is
important to determine if $\delta_g > \delta_{VH}$ or not.

The theoretical situation is less clear if the antiferromagnetic channel is
dominant.
There is no generic small parameter to control fluctuations in this
regime.
However, if the antiferromagnetic channel is dominant and the Fermi surface is
nearly nested, a controlled calculation turned out to be possible.
In this case we showed that the physics is controlled by classical spin
fluctuations with correlations which grow exponentially as $T$ decreases
below a mean field scale $T_{AF}$.
These quasistatic fluctuations produce a gap, $\Delta_F$, in the fermion
spectrum near the van Hove points which varies as
$\Delta_F = T_{AF} \sqrt{\frac{T_{AF}-T}{T_{AF}}}$ for $T<T_{AF}$.
This mechanism for producing a pseudogap predicts a very rapid $T$-dependence
of NMR rates which is not observed in experiment on superconducting materials
\cite{NMR}.

We now discuss the doping dependence of our results.
In a rigid band model one would expect $T_p$ to rise and then fall as the
chemical potential is tuned up to the van Hove point and then beyond.
However, in the spin model the particle-hole symmetry breaking
term $t'$ itself depends on doping as seen in Eq. (\ref{t'}).
As the doping is decreased beyond the van Hove point the Fermi surface gets
flatter.
This tends to pin the Fermi level to the corners.
Further, as the Fermi surface flattens, $p_0$ increases, so that $\omega_0$ in
Eq. (\ref{omega_0}) decreases and $T_p$ increases.
However, $\omega_\alpha$ also decreases so the importance of antiferromagnetism
grows.
As the doping is reduced towards zero, all instability scales become greater
than both $\omega_0$ and $\omega_\alpha$.
In this regime the pairing and antiferromagnetic charges scale in the same
way, as shown in Fig. 3.
Because we expect the bare value of the antiferromagnetic charge to be larger
than the bare value of the pairing charge we expect antiferromagnetism to be
dominant, as was also found by Dzyaloshinshkii and Yakovenko
\cite{Dzyaloshinskii87}.

Within our leading logarithm approximation we have shown that the observed
pseudogap can not be due to long-ranged antiferromagnetic fluctuations, and
must be due to RVB pairing.
We now offer qualitative arguments that this conclusion survives even the
leading logarithm approximation is not reasonable.
First, it seems that antiferromagnetic fluctuations can produce a pseudogap
only if they are long ranged and quasistatic; such fluctuations are ruled out
by NMR so we believe that a pairing origin of a pseudogap is more likely.
Second, increases in the antiferromagnetic charge seem to feed back into the
pairing equation in a way that increases $T_c$, so it seems natural to expect
that the theory with antiferromagnetic charge $\sim 1$ is unstable to $d$-wave
pairing {\em if} the Fermi surface is not too flat.
However, when the Fermi surface becomes flat, antiferromagnetism becomes
favored.
On the basis of these considerations we propose the phase diagram shown in
Fig. 4.

\vspace{.25in}
\centerline{\epsfxsize=8cm \epsfbox{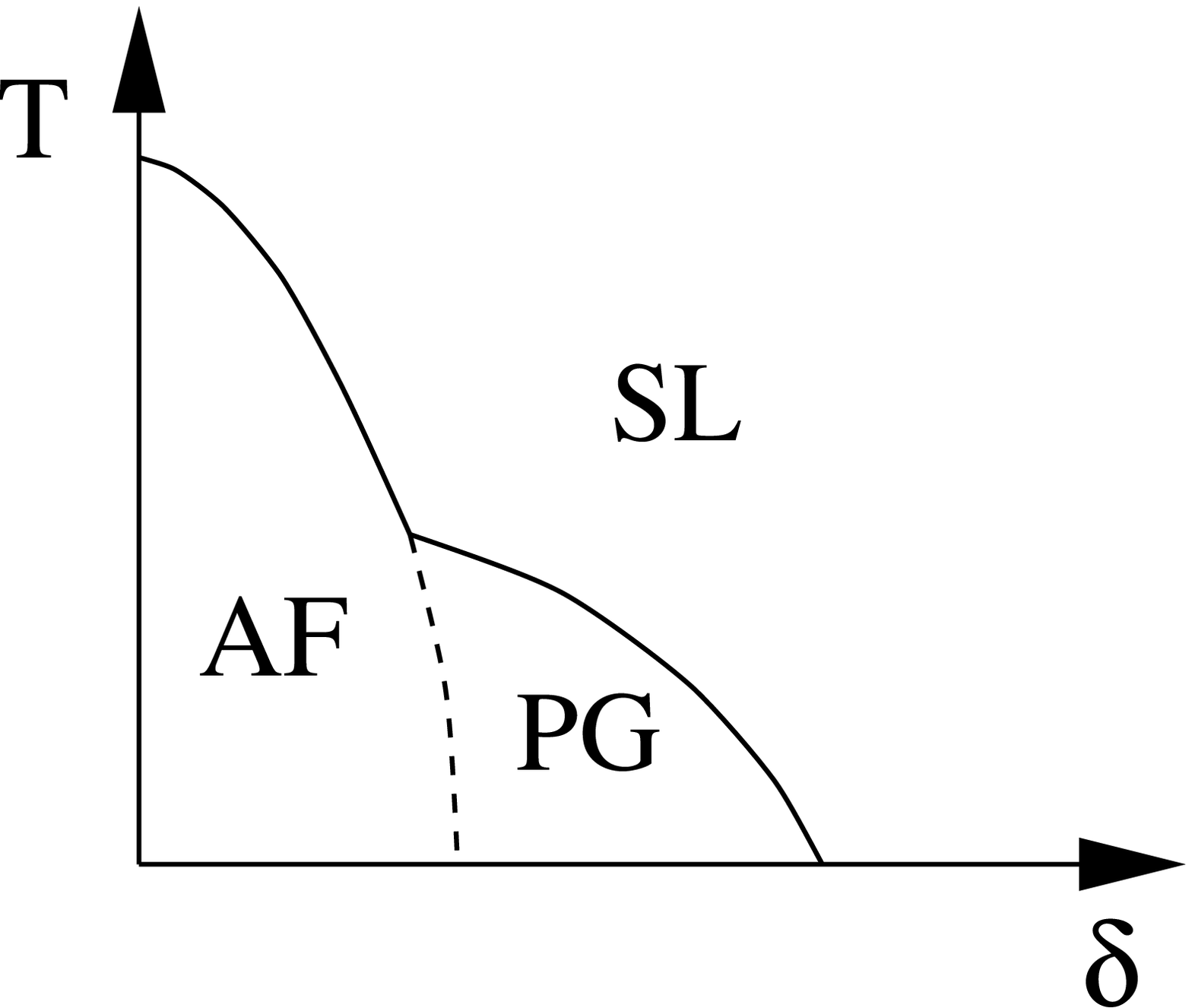}}

{\small
Fig 4.
Phase diagram of the spin liquid near van Hove point. AF denotes
antiferromagnetism, SL denotes spin liquid with ungapped Fermi surface and PG
denotes the pseudogap regime of the spin liquid phase}
\vspace{.1in}

Finally, we discuss the spin fluctuations expected in a different regime of
this phase diagram.
In the large $\delta$, spin liquid, regime and in the pseudogap regime near
the SL-PG boundary we expect spin fluctuations dominated by the ``$2p_F$''
effects discussed elsewhere \cite{Altshuler95a}.
As $\delta$ is decreased we expect enhancement of the spin fluctuations near
$(\pi,\pi)$ point due to the flattening of the Fermi surface and the increase
of the renormalized interaction.
Moreover, inside the PG regime there will be a gap on those parts of the Fermi
surface which could damp a $(\pi,\pi)$ fluctuation.
Therefore these fluctuations would be undamped as is apparently required by
analyses of $Cu$ NMR $T_1$ and $T_2$ experiments \cite{Sokol93}.

In summary, we have studied the effect of van Hove points on the ``uniform
RVB'' state with weak residual interactions.
We have identified three instabilities: to a flux phase with spontaneous
breaking of time reversal symmetry, magnetic phase, and a d-wave RVB phase.
We argued that the d-wave state is likely to happen at large doping while
antiferromagnetism dominates at small doping.
Our proposed phase diagram is shown in Fig. 3.
We showed that the inelastic scattering due to the gauge field is negligible
for fermions near the corners and that this in combination with the divergent
density of states permits a continuous crossover to a d-wave RVB state,
which is not allowed in models without corners.

\begin{appendix}

\section{Parameters of the Spinon Hamiltonian}

The spinon Hamiltonian is obtained expanding about a mean field approximation
to the $t-J$ model which may be written
\begin{equation}
H_{tJ} = - \sum_{i,j} t^{(band)}_{ij} b_i^\dagger b_j
	c_{j\alpha}^\dagger c_{i\alpha} + \frac{J}{2} \sum \vec{S_i} \vec{S_j}
\label{H_tJ}
\end{equation}
Here $c^\dagger$ is the spinon creation operator discussed in the text,
$\vec{S}_i=\frac{1}{2} \sum_{\alpha \beta} c_{i\alpha}^\dagger
\vec{\sigma}_{\alpha \beta} c_{i\alpha}$, the
bose operator $b_i^\dagger$ creates
a spinless charge $1$ ``holon'' and the constraint $b_i^\dagger b_i +
\sum_\alpha  c_{i\alpha}^\dagger c_{i\alpha}=1$ is assumed.
To obtain the fermion dispersion one  writes the
second term in $H_{tJ}$ as a product of four fermion operators \cite{Ioffe89}
\begin{equation}
\vec{S_i}\vec{S_j} = - \frac{1}{2} c_{j\alpha}^\dagger c_{i\alpha}
	c_{i\beta}^\dagger c_{j \beta} + \frac{1}{4}
\label{SS}
\end{equation}
and approximates this by
\begin{equation}
\vec{S_i}{S_j} \approx - \frac{1}{2} (c_{j\alpha}^\dagger c_{i\alpha} + h.c.)
\langle c_{i\beta}^\dagger c_{j \beta} \rangle .
\label{SS_app}
\end{equation}
If $\langle c_{i\beta}^\dagger c_{j \beta} \rangle \neq 0$ then the bosons
acquire a dispersion $H_{bose}=-\langle c_{i\beta}^\dagger c_{j \beta} \rangle
t_{ij}^{band} b_i^\dagger b_j$; because the density of bosons is low the
bosons are mostly in small $k$ states so it follows that for nearby $i$ and
$j$ $\langle b_i^\dagger b_j \rangle =\delta$.
Inserting this in Eq. (\ref{H_tJ}) we get the fermionic part of Eq. (\ref{H})
with
\begin{equation}
t_{ij} = \delta t_{ij}^{(band)} + \frac{1}{2} J
\langle c_{i\beta}^\dagger c_{j \beta} \rangle
\label{t_ij}
\end{equation}
The expectation value $\langle c_{i\beta}^\dagger c_{j \beta} \rangle$ turns
out to be $\frac{4}{\pi^2} + O(\delta)$.

We now estimate the bare gauge stiffness $g_0$.
This represents the effect of short wavelength fluctuations which are
integrated out in the definition of Eq. (\ref{H}); in the $N \rightarrow
\infty$ limit $g_0^{-2}$ vanishes \cite{Ioffe89}, but it is non-zero for the
physical $N=2$.
If the only term in $H$, Eq. (\ref{H}), were $\frac{1}{4g_0^2} f_{\mu\nu}^2$,
then one would find
\begin{equation}
\int_{-\infty}^{\infty} dt \langle h_i(t) h_i(0) \rangle = g_0^2
\end{equation}
Here as usual $h=\partial_xa_y-\partial_ya_x$.
To find $g_0^2$ we evaluate the $\langle h_i(t) h_i(0) \rangle$ correlator
directly from $S=1/2$ Heisenberg model, including in our calculation only
short wavelength degrees of freedom.
Formally, the gauge field is related to the phase of the operator
$\Delta_{ij}$ which decouples the four fermion operator in Eq. (\ref{SS})
\cite{Ioffe89}; the field $h$ is equal to the gauge flux through the
elementary placquette with vertices $i,j,k,l$; the latter is related to the
$\Delta$ operators by
\begin{equation}
R e^{ih} = \Delta_{ij} \Delta_{jk} \Delta_{kl} \Delta_{li}
\label{Re^h_1}
\end{equation}
Here $R$ is a real operator.
By undoing the decoupling one may express the product of four $\Delta$ around
the placquette in terms of fermion operators; in the low doping limit these in
turn may be expressed in terms of spin operators (with corrections
$O(\delta)$), yielding
\begin{equation}
\begin{array}{rl}
R e^{ih} = &
\frac{1}{8} + \frac{1}{2}\sum_{(ij)} \vec{S}_i\vec{S}_j \\
+ & i \sum_{(ijk)} \vec{S}_i(\vec{S}_j \times \vec{S}_k)  \\
	+ 2 [ (\vec{S}_i\vec{S}_j)(\vec{S}_k\vec{S}_l) + \! & \!\!
	(\vec{S}_i\vec{S}_l)(\vec{S}_k\vec{S}_j) \! -\!
	(\vec{S}_i\vec{S}_k)(\vec{S}_j\vec{S}_l)  ]
\end{array}
\label{Re^h_2}
\end{equation}
where $(ij)$ denotes all distinct pairs of spins, $(ijk)$ denotes all distinct
triads with the clockwise order of $i,j,k$ around the placquette.
If there are strong short-range antiferromagnetic correlations then we may
replace the real terms by c-numbers, obtaining
\begin{equation}
R e^{ih} = \frac{1}{2} + i \sum_{(ijk)} \vec{S}_i(\vec{S}_j \times \vec{S}_k)
\label{Re^h_3}
\end{equation}
This allows us to identify
\begin{equation}
h = 2  \sum_{(ijk)} \vec{S}_i(\vec{S}_j \times \vec{S}_k)
\label{h}
\end{equation}
and thus to relate $g_0^2$ to a six spin correlator.
Evaluating this in the spin wave approximation leads to
\begin{equation}
g_0^2 = B \sum_{k,p} \frac{1}{\omega_k+\omega_p}
\label{g_0^2}
\end{equation}
where $B$ is a number of the order of unity and $\omega_k$ and $\omega_p$ are
spin wave energies.
Because the typical value for $\omega_k \sim 2J$
and the sums are dominated by short wavelengths,
we see that $g_0^{-2} = B'
J$ with $B'$ a number of the order of unity.
Of course we can not expect this simple estimate to yield a reliable value
for $B'$, but it does show that it is reasonable to expect a large
contribution to the gauge stiffness from short range correlations.

Finally we consider the effective between planes hopping $t_\perp$.
If $t_\perp \neq 0$ then also the between planes exchange $J_\perp \neq 0$ and
we must add to Eq. (\ref{H_tJ}) the terms
\begin{equation}
H_\perp = \sum_i t_\perp b_i^{(1)\dagger} b_i^{(2)}
	c_{i\alpha}^{(1)\dagger} c_{i\alpha}^{(2)} +
	\frac{J_\perp}{2} \sum \vec{S_i}^{(1)} \vec{S_j}^{(2)}
\label{H_perp}
\end{equation}

One may factorize the four fermion term as in Eqs. (\ref{SS},\ref{SS_app});
however the equation determining the amplitude is
\begin{eqnarray}
\Delta_\perp &\equiv& (J_\perp + \delta t_\perp)
	\langle c_{i\beta}^{(1)\dagger} c_{i \beta}^{(2)} \rangle
\nonumber \\
	&=& \sum_{\epsilon,p} \frac{(J_\perp + \delta t_\perp) }
	{(i\epsilon-\epsilon_p)^2-\Delta_\perp^2}
\label{Delta_perp}
\end{eqnarray}
A solution only becomes possible for $J_\perp+\delta t_\perp \approx t$, so
for physically relevant parameters there is no coherent between planes hopping
in the model.

\end{appendix}

\end{multicols}

\begin{references}
\bibitem{Baskaran87} G. Baskaran, Z. Zou and P. W. Anderson,
Solid State Comm. {\bf 63} 973 (1987).

\bibitem{Ioffe89} L. B. Ioffe and A. I. Larkin, Phys. Rev. B {\bf 39} 8988
(1989).

\bibitem{Lee90} P. A. Lee, p. 96 in {\em High Temperature Superconductivity:
Proceedings}, K. S. Bedell, D. Coffey, D. E. Meltzer, D. Pines and
J. R. Schreiffer, Addison Wesley (Reading, MA: 1990);
L. B. Ioffe and B. G. Kotliar, Phys. Rev. B {\bf 42} 10348 (1990);
L. B. Ioffe, V. Kalmeyer and P. B. Wiegmann, Phys. Rev. B {\bf 43} 1219
(1991).

\bibitem{Dessau95} D. Dessau et al., unpublished.

\bibitem{Schreiffer88} J. R. Schreiffer, X. G. Wen and S. C. Zhang,
Phys. Rev. Lett. {\bf 60}, 944 (1980);
A. Kampf and J. R. Schreiffer, Phys. Rev. {\bf B41}, 6399 (1990) and
Phys. Rev. B{\bf 42}, 7967 (1990).

\bibitem{Lee73} P. A. Lee, T. M. Rice and P. W. Anderson,
Phys. Rev. Lett. {\bf 31}, 462 (1973).

\bibitem{Sadovskii74} M. V. Sadovskii, Zh. Eksp. Teor. Fiz. {\bf 66} 1720
(1974).

\bibitem{NMR} See, e.g. A. J. Millis, p. 198 in {\it High Temperature
Superconductivity:  Proceedings}, eds. K. S. Bedell, D. Coffey, D. E. Meltzer,
D. Pines and J. R. Schrieffer, (Addison Wesley:  Redwood, CA) 1990
or C. P. Slichter, p. 427 in {\it Strongly Correlated Electronic Materials},
eds. K. S. Bedell, Z. Wang, D. Meltzer, A. V. Balatsky and E. Abrahams,
Addison Wesley (Redwood City, CA)  1990.

\bibitem{Anderson87} P. W. Anderson, Science {\bf 235} 1196 (1987).

\bibitem{Affleck88} I. A. Affleck and J. B. Marston, Phys. Rev. B {\bf 37},
3774 (1988).

\bibitem{Varma88}C. M. Varma, S. Schmitt-Rink and E. Abrahams,
Sol. St. Comm. {\bf 62} 681 (1987).

\bibitem{Altshuler95a} B. Altshuler, L. B. Ioffe, A. I. Larkin and A. J. Millis
Phys. Rev. B {\bf 52}, 4607 (1995).

\bibitem{Altshuler95b} B. Altshuler, L. B. Ioffe and A. J. Millis, Phys. Rev. B
{\bf 52}, 5563 (1995).

\bibitem{Spielman90} S. Spielman, K. Fester, C. B. Eom, T. H. Geballe, M. M.
Fejer and A. Kapitulnik, Phys. Rev. Lett. {\bf 65}, 123 (1990).

\bibitem{Tanamoto92} T. Tanamoto, K. Kohno and H. Fukuyama, J. Phys. Soc. Jpn.
{\bf 61} 1886 (1992).

\bibitem{Ubbens94} M. Ubbens and P. A. Lee, Phys. B {\bf 50}, 438 (1994).

\bibitem{Wen95} X-G. Wen and P. A. Lee, cond-mat preprint/9506065.

\bibitem{Lee89} P. A. Lee and N. Read, Phys. Rev. Lett. {\bf 50}, 2691 (1987).

\bibitem{Dagotto95} E. Dagotto, A. Nazarenko and
A. Moreo, Phys. Rev. Lett. {\bf 74}, 310 (1995).

\bibitem{Newns92} D. C. Newns, C. C. Tseui, P. C. Pattnaik and C. L. Kane,
Comm Cond Mat. Phys. {\bf 15}, 273 (1992); D. M. Newns and P. C. Paittnaik p.
146 in {\it Strong Correlations and Superconductivity} eds. H. Fukuyama,
S. Maekawa and A. Malozemoff, (Springer-Verlag, Heidelberg: 1989).

\bibitem{Dzyaloshinskii87} I. E.  Dzyaloshinshkii,  Sov. Phys. JETP {\bf 66},
848 (1987);
I. E.  Dzyaloshinshkii and V. Yakovenko, Sov. Phys. JETP {\bf 67}, 844 (1988);
Int. J. Mod. Phys. B {\bf 2}, 667 (1988).

\bibitem{Pickett89} See, e.g. Table I of W. E. Pickett, Rev. Mod. Phys. {\bf
61}, 463 (1989).

\bibitem{Timusk95} See, e.g. the tabulation of observed penetration depths in
C. C. Homes, T. Timusk, D. A. Bonn, R. Liang and W. N. Hardy, cond/mat9509128.

\bibitem{Altshuler94} B. L. Altshuler, L. B. Ioffe and A. J. Millis,
Phys. Rev. B {\bf 50} 14048 (1994).

\bibitem{Polchinski94} J. Polchinski, Nucl. Phys. B, {\bf 422}, 617 (1994).

\bibitem{Landau-Lifshitz} L. D. Landau and E. M. Lifshitz, {\em Statistical
Physics}, vol. II, \S45.

\bibitem{Hwang94} H. Y. Hwang, B. Batlogg, H. Takagi, H. L. Kao, J. Kwo,
R. J. Cava, J. J. Krajewski and W. F. Peck, Jr, Phys. Rev. Lett. {\bf 72},
2636 (1994).

\bibitem{Barzykin94} V. Barzykin, D. Pines, A. V. Sokol and D. Thelen,
Phys. Rev. B {\bf 49} 1544 (1994).

\bibitem{Altshuler95c} B. Altshuler, L. B. Ioffe and A. J. Millis,
Phys. Rev. B, to be published.

\bibitem{Sokol93} A. V. Sokol and D. Pines, Phys. Rev. Lett. {\bf 71} 2813
(1993).


\end{references}
\end{document}